\documentclass{nature_withfigs}
\pdfoutput=1

\usepackage{amsmath,amsfonts,amssymb}
\usepackage{graphicx}

\title{Interference of interacting matter waves}

\author{Mattias Gustavsson$^1$, Elmar Haller$^1$, Manfred J. Mark$^1$, Johann G.
Danzl$^1$,\\Russell Hart$^1$, Andrew J. Daley$^{2,3}$  \&  Hanns-Christoph N\"agerl$^1$}

\begin{document}

\maketitle

\begin{affiliations}
\item Institut f{\"u}r Experimentalphysik und Zentrum f\"{u}r Quantenphysik, Universit{\"a}t Innsbruck, \newline Technikerstra{\ss}e 25, A--6020 Innsbruck, Austria
\item Institut f\"ur Theoretische Physik und Zentrum f\"{u}r Quantenphysik, Universit{\"a}t Innsbruck, \newline Technikerstra{\ss}e 25, A--6020 Innsbruck, Austria
\item Institut f\"ur Quantenoptik und Quanteninformation der \"{O}sterreichischen Akademie der Wissenschaften, Technikerstra{\ss}e 21a, A--6020 Innsbruck, Austria
\end{affiliations}

\begin{abstract}
The phenomenon of matter wave interference lies at the heart of quantum physics.
It has been observed in various contexts in the limit of non-interacting particles as a single particle effect.
Here we observe and control matter wave interference whose evolution is driven by interparticle interactions. In a multi-path matter wave interferometer, the macroscopic many-body wave function of an interacting atomic Bose-Einstein condensate develops a regular interference pattern, allowing us to detect and directly visualize the effect of interaction-induced phase shifts. We demonstrate control over the phase evolution by inhibiting interaction-induced dephasing and by refocusing a dephased macroscopic matter wave in a spin-echo type experiment. Our results show that interactions in a many-body system lead to a surprisingly coherent evolution, possibly enabling narrow-band and high-brightness matter wave interferometers based on atom lasers.
\end{abstract}

Matter wave interference has been observed as a single particle effect for electrons\cite{Davisson1927}, neutrons\cite{Halban1936}, atoms and molecules\cite{Estermann1930}. Macroscopic matter wave interference was first directly observed in the case of two independent atomic Bose-Einstein condensates (BEC) that were brought to overlap\cite{Andrews1997}. This experiment validated the notion of the BEC as a macroscopic matter wave and coined the expression of the atom laser in analogy to the laser for the case of photons. Matter wave interferometers\cite{Berman1997,Cronin2007,Hart2007}, in particular for applications to precision measurements, are typically operated in the dilute single particle limit\cite{Wicht2002,Clade2006,Fixler2007} to avoid particle-particle interactions. Atom interferometers based on Bose-Einstein condensates (BEC) are expected to benefit from the extremely low momentum spread, the exceptional brightness, and the low spatial extent of the BEC\cite{Gupta2002}, but they readily enter the nonlinear matter wave regime as a result of the interaction-induced mean field potential. A possible solution is to operate BEC-based interferometers in the non-interacting limit\cite{Gustavsson2008a,Fattori2008} by exploiting the cancellation of the scattering phase shift near a scattering resonance. This condition, however, is difficult or impossible to fulfill for most atomic species. In the present work we demonstrate a BEC-based multipath atom interferometer where the dynamics is dominated by interaction-induced phase shifts. We realize the multipath interferometer by loading an interacting BEC into an optical lattice potential along one dimension, coherently splitting the BEC into several parts that are then each subject to different linear and nonlinear phase shifts. The linear phase shifts due to the gravitational force lead to the the well-known phenomenon of Bloch oscillations\cite{Dahan1996,Anderson1998}, whereas the interaction-induced nonlinear phase shifts cause the macroscopic wave function to first spread in momentum space as a function of time and then, surprisingly, to exhibit high-contrast interference. We demonstrate a high degree of coherence by reversing the nonlinear phase evolution, thereby refocusing the BEC momentum wave function. By application of an external potential we cancel the dominant mean-field contribution to the phase evolution and become sensitive to beyond-mean-field effects. A crucial ingredient of our experiments is the capability to tune $a$, the atomic scattering length which determines the strength of the interaction, by means of a Feshbach resonance\cite{Weber2003}. In particular, $a$ can be switched to zero to stop the interaction driven part of the evolution in the interferometer or to perform high resolution wave function imaging in momentum space.

\section{Phase evolution in the matter wave interferometer}

Our interferometer consists of a BEC in adiabatically loaded into a 1D optical lattice potential with a superimposed harmonic trap, as illustrated in Fig. 1a. In the tight-binding regime, it is convenient to write the macroscopic wave function of the condensate, $\Psi$, in a basis\cite{Smerzi2003} of wave functions $\Psi_j(z,r_\perp)$ centered at the position $z_j=jd$ of the individual lattice sites $j$, $\Psi (z, r_\perp, t) = \sum_j c_j(t) \Psi_j(z,r_\perp)$. Here, $z$ is the coordinate along the (vertical) lattice direction, $ r_\perp$ is the transverse coordinate, $d$ is the distant between adjacent lattice sites and $c_j(t)$ are time-dependent complex amplitudes.

After the BEC is loaded into the lattice, we tilt the lattice potential by applying a strong force $F$ along the lattice direction. In the limit $Fd \gg J$, where $J$ is the tunneling matrix element, tunneling between lattice sites is inhibited. The on-site occupation numbers $|c_j|^2$ are then fixed, and we can write $c_j(t) = c_j(0) e^{i\phi_j(t)}$, where the phase $\phi_j(t)$ evolves in time according to the local potential at each specific lattice site\cite{Witthaut2005},
\begin{align}
	\hbar \frac{\partial \phi_j}{\partial t} &= Fdj + V_j^{\rm trap} + \mu^{\rm loc}_j \nonumber \\
	&= Fdj + \beta_{\rm tr} j^2 - \alpha_{\rm int} j^2 .
\end{align}
Here, the total potential at each lattice site $j$ consists of three terms. The applied force leads to a term linear in $j$ and causes Bloch oscillations\cite{Dahan1996,Anderson1998} with angular frequency $Fd/\hbar$. The second term comes from an optional harmonic confinement, where $\beta_{\rm tr}=m \omega_{\rm tr}^2 d^2/2$ characterizes the strength of the confining potential and $\omega_{\rm tr}$ is the trap frequency. Atom-atom interactions give rise to a third term, the local chemical potential $\mu^{\rm loc}_j$, which depends on the scattering length $a$ and the site occupation number as\cite{Smerzi2003} $\mu^{\rm loc}_j \propto \sqrt{a |c_j|^2}$. When the BEC is loaded in the Thomas-Fermi regime, as is done here, its initial value can be calculated in a simple way. The density distribution will be such that the local chemical potential mirrors the trapping potential used when loading into the lattice, $\mu^j_{\rm loc} = \mu - V_j^{\rm trap}$, with $\mu$ being the (global) chemical potential of the BEC. We then initially have $\mu^{\rm loc}_j = \alpha_{\rm int} j^2$, where $\alpha_{\rm int}=m \omega_{\rm lo}^2 d^2/2$ and $\omega_{\rm lo}$ is the trap frequency during loading.

The phase terms proportional to $j^2$ lead to a nonlinear phase evolution and to a dephasing of the lattice site. This results in a time-varying interference pattern of the macroscopic matter wave, as we will demonstrate below. The key in our experiments is that we have full control over these nonlinear terms, not only over $\beta_{\rm tr}$ via the external trapping potential, but also over the interaction term characterized by $\alpha_{\rm int}$, both via the initial density distribution, and, more importantly, via the scattering length $a$. By tuning the scattering length\cite{Weber2003} from its initial value $a$ to $a^\prime$, we can ramp $\alpha_{\rm int}$ to a new value $ \alpha_{\rm int}^\prime$, which can in particular be set to zero for $a\!=\!0$. Nonlinear phase terms for matter waves are well known in single particle quantum mechanics. They play an important role for matter wave Talbot interferences\cite{Berman1997,Deng1999} and can be visualized in terms of so-called matter wave quantum carpets\cite{Kaplan2000}. In these contexts, the phase terms arise from propagation. In our case, the nonlinear phase terms for $\alpha_{\rm int}\!\ne\!0$ arise from interactions and thus lead to a density dependent many-body effect in the multipath atom interferometer.

In the preceding discussion, we have assumed that the minimum of the trapping potential is centered directly over one of the lattice minima. If this is not the case, the trapping potential term in equation (1) has to be modified to $\beta_{\rm tr}(j-\delta)^2 = \beta_{\rm tr} j^2 - 2 \beta_{\rm tr} \delta j + \textit{const.}$, where $\delta \in [0,1]$ describes the offset of the trap center in the $z$-direction with respect to the lattice minima, and an analogous modification has to be done to the interaction term. This adds a small term linear in $j$ and therefore leads to a slight modification of the Bloch oscillation frequency. In our experiments, $\delta$ is the only parameter that we do not fully control. It is constant on the timescale of a single experimental run, but it drifts over the course of minutes as the beam pointing of the horizontally propagating laser beam generating the trapping potential is not actively stabilized.

\section{Interaction-induced matter wave interference}

The starting point for our experiments is a BEC trapped in a crossed optical dipole trap and adiabatically loaded into an optical lattice, as illustrated in Fig. 1a. The gravitational force acting on the BEC is initially compensated using magnetic levitation\cite{Weber2003}. We effectively start the multipath atom interferometer and hence the evolution of the interacting macroscopic wave function by turning off magnetic levitation and ramping down the vertical confinement created by laser beam L$_2$ within 0.3 ms, inducing Bloch oscillations in the lowest band of the lattice. With $ Fd/\hbar \approx 2 \pi \times 1740$ Hz and $ J/\hbar \approx 2 \pi \times 40$ Hz the on-site occupation numbers $|c_j|^2$ are fixed to their initial values. After an evolution time $\tau$, we close the interferometer by ramping down the lattice in 1 ms and directly image the (vertical) quasi-momentum distribution in the first Brillouin zone (BZ). The ramp is adiabatic with respect to the bandgap and maps quasi-momentum onto real momentum\cite{Kastberg1995}, which is measured by taking an absorption image after a period of free expansion.
Fig. 1b shows absorption images of the first Bloch oscillation\cite{Dahan1996}. The Bloch period is about 0.58 ms and the peaks have a root mean square (rms) width of $0.2\hbar k$, where $k=\pi/d$ is the lattice wave vector, thus being well separated.

We study the evolution of the wave function at high resolution in momentum space by taking snapshots after extended time-of-flight.
As illustrated in Fig. 2a, the BEC wave function spreads out in the BZ in about $N=18$ Bloch cycles. Then, surprisingly, an interference pattern gradually develops at the edge of the BZ and later also becomes visible at the center of the BZ, while the number of interference maxima and minima changes as time progresses. Images are taken after an integer number of Bloch cycles for cycle phase $\phi=0$, corresponding to the first image in Fig. 1b. The data is acquired with an interacting BEC with the scattering length set to 190 a$_0$, where a$_0$ is the Bohr radius, at an initial peak density of $n = 4 \times 10^{13}$ atoms/cm$^3$, occupying about 35 lattice sites after loading. We can follow the evolution of the interference pattern for more than $N=100$ Bloch cycles, corresponding to times beyond 60 ms. This is about a factor 10 longer than the timescale for the initial broadening. We find that the number of maxima and minima and their location in the interference pattern as measured after fixed evolution time $\tau$ depend on the initial atomic density, on the strength of the interaction, and on the number of occupied lattice sites.
We also find that the measured quasi-momentum distribution for a given $\tau$ is reproducible from one experimental realization to the next, except that the pattern appears slightly shifted within the BZ after several experimental realizations. We attribute this to a drift of $\delta$, the offset of the lattice minima from the dipole trap center, which leads to a small change of the Bloch frequency as noted before. We do not actively stabilize the vertical position of L$_2$ with respect to the lattice, and hence temperature variations in the laboratory slowly change $\delta$.

We combine two techniques to achieve a high resolution in momentum space and to visualize the interference pattern. First, we minimize broadening of the distribution as a result of interactions by setting $a$ to zero during the release from the lattice and the subsequent free expansion\cite{Gustavsson2008a}. In addition, we use long expansion times, employing magnetic levitation to prevent the BEC being accelerated by gravity and falling out of the field of view. Fig. 3 shows how the contrast emerges during the expansion for a BEC after $N\!=\!40$ Bloch cycles. It takes more than 100 ms of expansion for the interference pattern to acquire full contrast. In general, we find that the contrast is improved when the horizontally confining beam L$_1$ is not switched off abruptly but is ramped down slowly within the first 55 ms of time-of-flight, reducing the horizontal expansion rate. However, this happens at the cost of some additional momentum broadening along the vertical direction. Our imaging techniques allow us to resolve structure in momentum space on a scale below $ 0.1 \hbar k $ in a single shot absorption image.

To understand the interference structure and its evolution in time, we compute the total BEC wave function in quasi-momentum space for a phase evolution at the different lattice sites given by Equation (1), as detailed in the Methods section. Fig. 2b shows the interference pattern for our experimental parameters according to this simple model. The experimental results are qualitatively  very well reproduced by the model when we reduce $\alpha_{\rm int}$ by a factor of 0.9 compared to the value deduced from our experimental parameters. This scale factor accounts primarily for the fact that our simple model does not take into account any horizontal dynamics. In particular, switching off L$_2$ when starting the evolution leads to an excitation of a radial breathing mode in the horizontal plane, reducing the density at each site and modulating it in time. To a first approximation, rescaling of $\alpha_{\rm int}$ accounts for this. Nevertheless, the agreement between the experiment and the analytical model indicates that the dominant driving mechanism for the wave function spreading and interference is the nonlinear phase evolution. In particular, phase coherence is not lost, in contrast to previous experiments\cite{Morsch2003}. We test this coherence and demonstrate control over the phase evolution in two experiments.

\section{Cancellation of dephasing using an external potential}

Equation (1) suggests that the effect of interactions can be cancelled by the application of an external potential. Indeed, choosing this potential to be equal to the initial loading potential, i.e. choosing $\alpha_{int} \approx \beta_{tr}$, allows us to observe persistent Bloch oscillations for an interacting BEC. The BEC quasi-momentum distribution after $N=40$ Bloch cycles is shown in Fig. 4a and 4b as a function of the strength of the external compensating potential, given by the power in laser L$_2$. When the external potential does not compensate for interactions, the condensate wave function is dephased and spreads over the whole BZ within less than $N=20$ Bloch cycles. In contrast, when the external potential balances the effect of interactions, the BEC wave function does not spread out and Bloch oscillations are clearly visible. The time during which Bloch oscillations can be observed is now greatly extended compared to the case when a compensating potential is absent.
The transition from a dephased to a non-dephased wave function as a function of confinement strength is quantified in Fig. 4c, where the rms-width $\Delta p$ of the singly-peaked quasi-momentum distribution after $N=40$ Bloch cycles is plotted as a function of the laser power in L$_2$.
Fig. 4d and 4e show the time evolution of the quasi-momentum distribution without and with the compensating potential while all other parameters are kept the same. Fig. 4d essentially shows the broadening of the distribution as described before. Interestingly, the condensate wave function in the presence of a compensating potential shown in Fig. 4e dephases in a completely different way. Initially, the central peak shows no broadening. However, it is slowly depopulated, while a much broader background distribution is increasingly populated. After about 100 oscillations, the shape of the central peak starts to develop side lobes or splits in two, with the exact shape varying from one experimental run to the next. The timescale for the loss of interference is a factor 10 larger than the timescale on which the dephasing and hence the initial broadening takes place in the uncompensated case.

\section{Rephasing of a dephased condensate}

Second, we perform a matter wave spin-echo-type experiment. We initially proceed as shown in Fig. 2, letting the wave function evolve for a time corresponding to about $N=40$ Bloch cycles until it is fully dephased and shows, upon measurement, a regular interference structure. We then essentially remove the effect of interactions by ramping to $a\!=\!10$ a$_0$ within $10$ ms. By not switching the interaction entirely off and by ramping comparatively slowly we avoid excessive excitation of the radial breathing mode as a result of the change in the mean field potential at each site. At the same time, we gradually turn on the harmonic potential as given by the horizontal dipole trapping laser beam L$_2$ within 4 ms to approximately the same depth as during the initial BEC loading phase. From equation (2) we expect that the wave function now experiences a phase shift with a quadratic spatial dependence with opposite sign, allowing us to reverse the evolution and to recover the initial condition. Fig. 5 shows the resulting quasi-momentum distributions. As time progresses, the wave function indeed refocuses while it continues to perform Bloch oscillations. As we do not control the value of $\delta$ for a particular run, we record about $10$ distributions for each evolution time and select those that are symmetrical, corresponding to Bloch cycle phase $\phi=0$ or $\phi=\pi$. For the chosen strength of the potential, refocusing happens after about 24 Bloch cycles after the ramp of $a$. This confirms that the initial broadening and dephasing mechanism must have been coherent. We note that we cannot avoid some excitation of the radial breathing mode as seen in the absorption images given in Fig. 5.

\section{Discussion}

Our results raise several important questions: To what extent can matter wave interferometry be performed in the presence of interactions? What sets the timescale for the eventual loss of interference contrast? Certainly, our simple analytic model does not predict any loss of contrast. In particular, it should be possible to completely eliminate the effect of interactions with the compensating external potential. However, there are several effects not included in the model that could cause the residual dephasing we observe. Motion in the radial direction, which causes the density and therefore the interaction energy to change over time, could lead to mixing of the different degrees of freedom and hence to additional dephasing. This might apply to our matter wave spin-echo experiment, but in the experiment where we compensate by means of the external potential there is hardly any radial excitation and this effect should not play a role. The appearance of dynamical instabilities\cite{Zheng2004,Cristiani2004,Fallani2004} can be ruled out, as our experimental parameters are outside the unstable region. Going beyond the mean-field treatment, a variety of factors can lead to dephasing. For example, at each lattice site there exists a superposition of number states, accumulating different phases corresponding to their respective interaction energies\cite{Tuchman2007,Imamoglu1997}. This leads to an effective dephasing, as the phase on a particular lattice site becomes ill-defined. Basic estimates\cite{Tuchman2007,Imamoglu1997} indicate a dephasing time of about 130 ms for our system, on the same order as we observe.

These experiments constitute a clear demonstration of coherent dynamics in an interacting macroscopic quantum system. This coherence affords a large degree of control over the system, as demonstrated by the possibility to rephase the wave function using an external potential in order to reverse dephasing due to interactions. The control demonstrated here has potential application in matter-wave interferometry, and such a degree of control over the mean-field evolution also opens the possibility to probe beyond-mean-field effects in atom interferometers.

\bigskip
\newpage
\noindent {\bf \large Methods}

\noindent {\bf Sample preparation}
\newline
Our experimental approach initially follows the procedure described in ref.\cite{Gustavsson2008a}. In brief, within $10$~s we produce an essentially pure BEC with tunable interactions\cite{Weber2003} in the Thomas-Fermi limit with up to $1.5 \! \times \! 10^5$ Cs atoms. The BEC is trapped in a crossed-beam dipole trap generated by a vertically (L$_1$) and a more tightly focused horizontally (L$_2$) propagating laser beam. The BEC is cigar-shaped with the long axis oriented along the direction of L$_2$. The trap frequencies are $(\omega_x,\omega_y,\omega_z)\!=\!2 \pi \times (39,5,39) $ Hz, where $x$ denotes the horizontal direction perpendicular to $L_2$, $y$ is the axial direction along $L_2$, and $z$ is the vertical direction. We magnetically control the scattering length $a$ in the range between $0$\,a$_0$ and $300$\,a$_0$ with a resolution of about $0.1$\,a$_0$. For BEC production, we work at $a\!=\!210$ a$_0$, where three-body losses are minimized\cite{Kraemer2006}. Initially, we support the optical trapping by magnetic levitation against gravity\cite{Weber2003}. As shown in Fig. 1a we superimpose an optical lattice with $d\!=\!\lambda/2$ along the vertical direction, where $\lambda\!=\!1064.5 $ nm is the wavelength of the lattice light. To load the BEC into the lattice, we stiffen the horizontal confinement within 1 s, leading to trap frequencies of $2 \pi \times (41,13,39)$ Hz, and at the same time turn on the lattice potential exponentially to a depth of $8 E_R$. Here, $E_R=h^2/(2 m \lambda^2) = k_B \! \times \! 64 \, $nK is the photon recoil energy and $m$ the mass of the Cs atom. The BEC is thus gently loaded into the lattice, occupying about 25 to 35 lattice sites, with up to 7000 atoms at the central site.

\newpage
\noindent {\bf Derivation of the BEC wave function in momentum space}

Here, we outline the method used to calculate the images in Fig. 2b. Due to the comparatively small interaction energies in our system, the atoms are restricted to move in the lowest Bloch band and we can write the local wavefunction at lattice site $j$ as $\Psi_j(r_\perp,z) = w_0^{(j)}(z) \Phi_{\perp}(\rho_j,r_\perp)$, where $w_0^{(j)}(z)$ is the lowest-band Wannier function localized at the j-th site and $\Phi_{\perp}(n_j,r_\perp)$ is a radial wave function depending on the occupation number $n_j=|c_j|^2$ at each site\cite{Smerzi2003}.
We can then write the total time-dependent wave function in momentum space as
\begin{align}
		\Psi(p_z,p_\perp,t) &= \sum_j c_j(t) w_0^{(j)}(p_z) \Phi_\perp(n_j,p_\perp) \nonumber \\
		&= w_0^{(0)}(p_z) \sum_j c_j(t) e^{-ip_z jd} \Phi_\perp(n_j,p_\perp).
\end{align}
Transforming to quasi-momentum space and assuming that the phase at each lattice site evolves according to Equation (2), we can write\cite{Witthaut2005}
\begin{equation}
		\Psi(q_z,p_\perp,t) = \sum_j c_j(0) e^{-i(q + \frac{Ft}{\hbar}) jd} \, e^{-i(\beta_{\rm tr} j^2 - \alpha_{\rm int} j^2) t / \hbar} \Phi_\perp(n_j,p_\perp),
\end{equation}
where $q_z$ denotes the quasimomentum. The images in Fig. 2b show the BEC density distribution $|\Psi(q_z,p_\perp,t)|^2$ integrated along one radial direction, using a Thomas-Fermi wave function as radial wave function $\Phi_\perp(n_j,p_\perp)$.

We have compared the result in Fig. 2b with a numerical integration of the discrete nonlinear Schrödinger equation\cite{Smerzi2003}, which includes tunneling between lattice sites, and find essentially identical results, confirming that tunneling is inhibited.


\begin{addendum}

\item[Acknowledgements] We thank E. Arimondo, O. Morsch, W. Schleich, A. Smerzi, D. Witthaut and A. Buchleitner and his group for helpful discussions. We are indebted to R. Grimm for generous support and gratefully acknowledge funding by the Austrian Ministry of Science and Research (Bundesministerium f\"ur Wissenschaft und Forschung) and the Austrian Science Fund (Fonds zur F\"orderung der wissenschaftlichen Forschung) in form of a START prize grant and through SFB 15. R.H. is supported by a Marie Curie International Incoming Fellowship within the 7th European Community Framework Programme.

\item[Competing Interests] The authors declare that they have no competing financial interests.

\item[Correspondence] Correspondence and requests for materials should be addressed to H.-C. N. \newline(email:christoph.naegerl@uibk.ac.at).

\end{addendum}

\begin{figure}
\begin{center}
\includegraphics{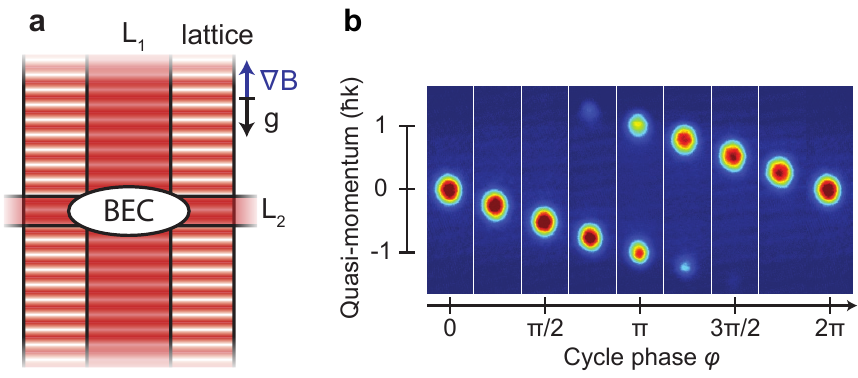}
\caption{{\bf BEC-based atom interferometer.} {\bf a,} Experimental configuration: The tunable BEC is formed at the intersection of the vertical guide laser beam L$_1$ and a horizontal trapping beam L$_2$. The lattice is oriented along the vertical direction. Gravity, g, is initially compensated by a force due to a magnetic field gradient, $\nabla B$. {\bf b,} Imaging the first Brillouin zone (BZ): One cycle of Bloch oscillations for a non-interacting BEC as seen in time-of-flight absorption imaging, showing narrow peaks cycling through quasi-momentum space for cycle phases $\phi\!=\!0$, $\pi/4$, $\pi/2$, ..., to $2\pi$.
}
\label{fig1}
\end{center}
\end{figure}

\begin{figure}
\begin{center}
\includegraphics{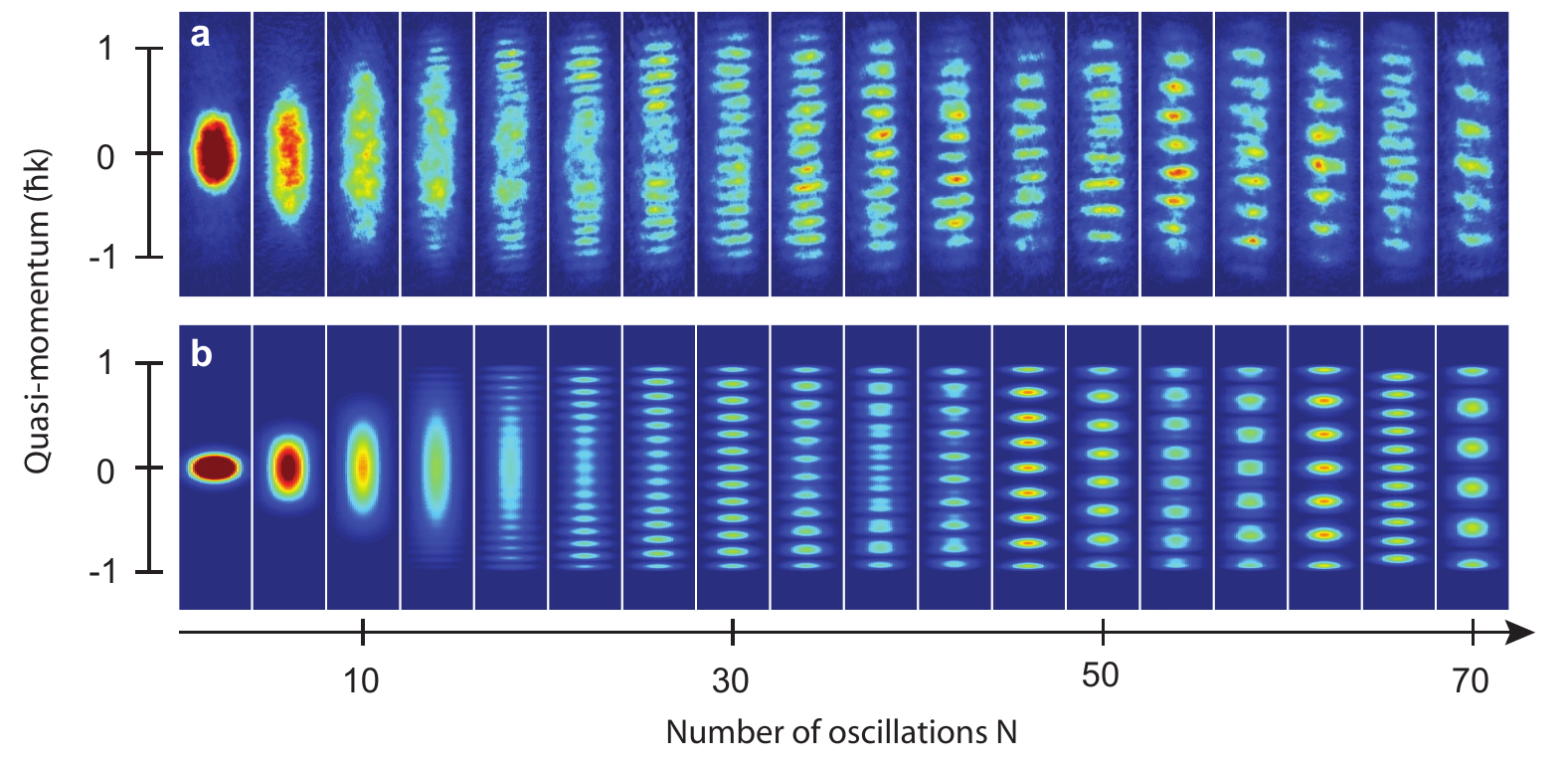}
\caption{{\bf Interaction induced macroscopic matter wave interference.} {\bf a,} Experimental results showing the quasi-momentum distribution as a function of evolution time $\tau$ given in units of the Bloch period. The absorption images are taken in steps of 4 Bloch cycles for a BEC with an initial peak density of $n\!=\!4 \times 10^{13} $ atoms/cm$^3$ loaded into about $35$ lattice sites with $ a\!=\!190$ a$_0$. Each image corresponds to a single realization of the experiment. {\bf b,} Evolution of the wave function in quasi-momentum space when the phase at the individual lattice sites evolves according to equation (2) with $\beta_{\rm tr}\!=\!0$ (no external trap) for $n\!=\!4 \times 10^{13} $ atoms/cm$^3$ loaded into $35$ lattice sites with $ a\!=\!190$ a$_0$.
$\alpha_{\rm int}$ is slightly rescaled to account for the reduction in density due to transversal dynamics, see text. In {\bf a}, some additional broadening, largely due to the presence of the horizontal trapping potential during expansion, can be seen.}
\label{fig2}
\end{center}
\end{figure}

\begin{figure}
\begin{center}
\includegraphics[width=9.5cm]{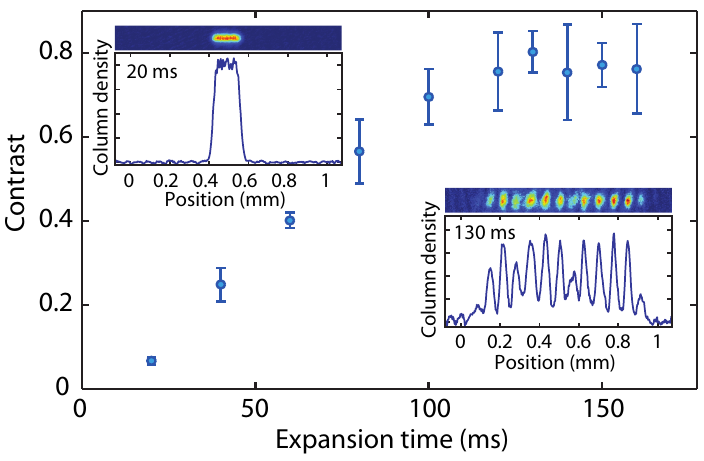}
\caption{{\bf Contrast of interference fringes.} Contrast of matter wave interference emerging during time-of-flight expansion for a BEC after $N\!=\!40$ Bloch cycles, where the wave function completely fills the BZ. We define the contrast as $(I_{max}-I_{min})/(I_{max}+I_{min})$, where $I_{max}$ ($I_{min}$) is the average value of the maxima (minima) of the central peak structure. Each data point is the average contrast of 10 experimental runs and the error bars indicate the $1 \sigma$ statistical error. The insets show measured quasi-momentum distributions integrated along the transverse direction at two expansion times as indicated.}
\label{fig3}
\end{center}
\end{figure}

\begin{figure}
\begin{center}
\includegraphics{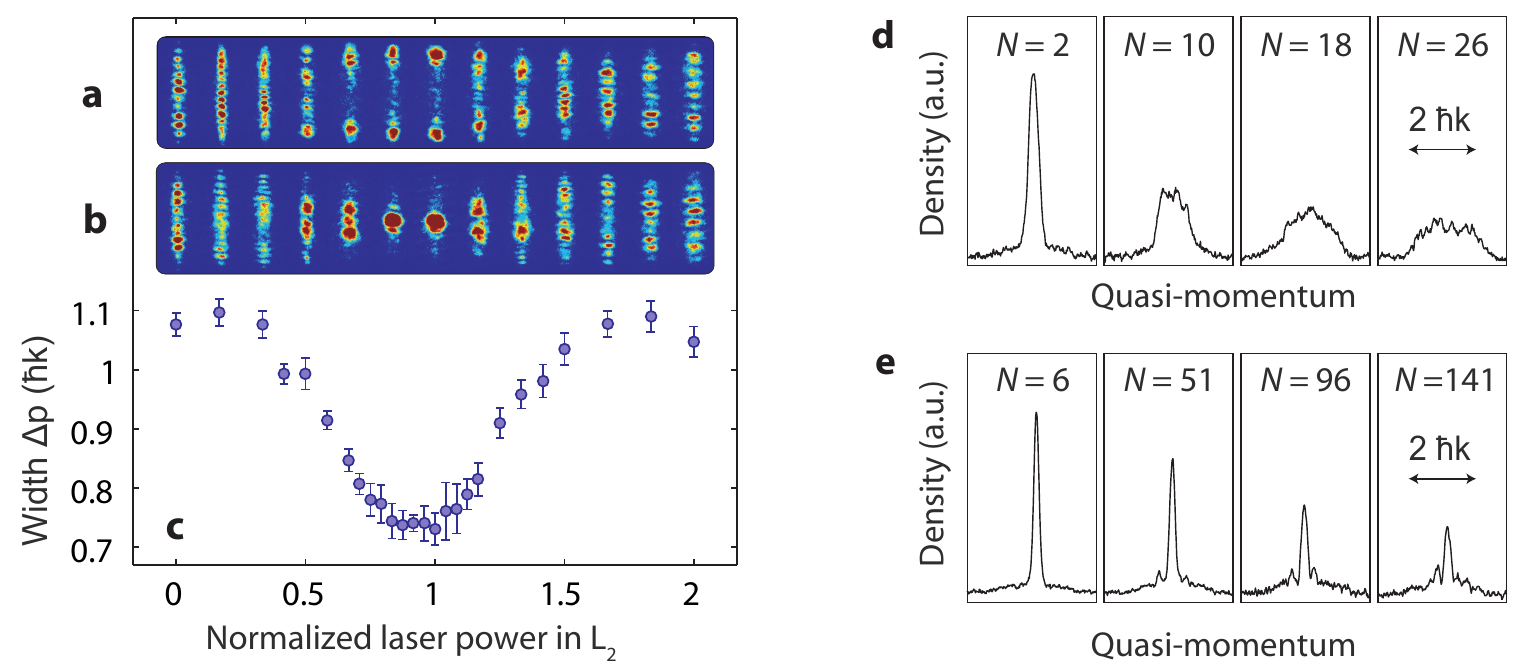}
\caption{{\bf Cancellation of interaction induced dephasing and observation of persistent Bloch oscillations.} {\bf a-c,} Absorption images showing the quasi-momentum distribution for cycle phase $\phi\!=\!\pi$ ({\bf a}) and $\phi\!=\!0$ ({\bf b}) after $N\!=\!40$ Bloch cycles and ({\bf c}) momentum width $\Delta p$ for $\phi\!=\!0$ as a function of confinement strength, normalized to the confinement strength at loading.
{\bf d} Momentum distribution for $\phi\!=\!0$ as a function of the number $N$ of Bloch cycles when no compensating potential is present, showing fast broadening. {\bf e} The evolution of the momentum distribution for the case of optimum cancellation of interactions.}
\label{fig4}
\end{center}
\end{figure}

\begin{figure}
\begin{center}
\includegraphics{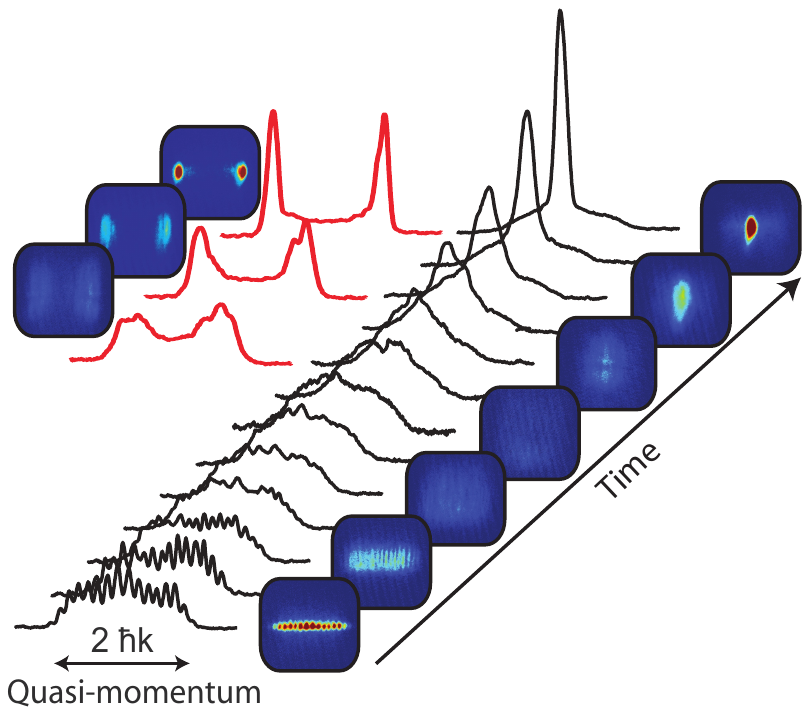}
\caption{{\bf Matter wave spin-echo-type experiment.} Rephasing of the BEC from a fully dephased wave function back into a narrow distribution after switching interactions to near zero and turning on an external potential. Time progresses from front to back. The black solid lines correspond to selected quasi-momentum distributions that refocus into the characteristic singly-peaked distribution (cycle phase $\phi\!=\!0$), see text. They are separated in time by 1.15 ms or two Bloch cycles, and they are offset for clarity. The red solid lines correspond to selected distributions that refocus into the characteristic double-peaked distribution (cycle phase $\phi\!=\!\pi$). The images are absorption images corresponding to the adjacent quasi-momentum distributions. Some radial excitation is also present.}
\label{fig5}
\end{center}
\end{figure}

\end{document}